\documentclass[aps,prd,floatfix,showpacs]{revtex4}
\usepackage[dvips]{graphicx}
\newcommand{\be}{\begin{equation}}
\newcommand{\ee}{\end{equation}}
\newcommand{\bea}{\begin{eqnarray}}
\newcommand{\eea}{\end{eqnarray}}
\newcommand{\bitem}{\begin{itemize}}
\newcommand{\eitem}{\end{itemize}}

\newcommand{\benum}{\begin{enumerate}}
\newcommand{\eenum}{\end{enumerate}}
\newcommand{\bc}{\begin{center}}
\newcommand{\ec}{\end{center}}
\begin{document}
\title{Scaling Symmetry and Integrable Spherical Hydrostatics}
\author{Sidney Bludman}
\email{sbludman@das.uchile.cl}
\homepage{http://www.das.uchile.cl/~sbludman}
\affiliation{Departamento de Astronom\'ia, Universidad de Chile, Santiago, Chile}
\author{Dallas C. Kennedy}
\email{dalet@stanfordalumni.org}
\homepage{http://home.earthlink.net/~dckennedy}
\date{\today}
\begin{abstract}
Any symmetry reduces a second-order differential equation to a first integral: variational symmetries of the action (exemplified by central field dynamics) lead to conservation laws, but
symmetries of only the equations of motion (exemplified by scale-invariant hydrostatics) yield first-order {\em non-conservation laws} between invariants. We obtain these non-conservation laws by extending Noether's Theorem to non-variational symmetries and present an innovative variational formulation of spherical adiabatic hydrostatics.  For the scale-invariant case, this novel synthesis of group theory, hydrostatics, and astrophysics allows us to 
recover all the known properties of polytropes and define a {\em core radius}, inside which polytropes of index $n$ share a common core mass density structure, and outside of which their envelopes differ. The Emden solutions (regular solutions of the Lane-Emden equation) are obtained, along with useful approximations. An appendix discusses
the $n=3$ polytrope in order to emphasize how the same mechanical structure allows different
thermal structures in relativistic degenerate white dwarfs and zero age main sequence stars.

\end{abstract}
\pacs{45.20.Jj, 45.50.-j,  95.30.Lz, 97.10.Cv}
\maketitle

\section{Symmetries of Differential Equations and Reduction of Order}

Noether's Theorem relates every {\em variational symmetry,} a symmetry of an action or similar integral, to a conservation law, a first integral of the equations of motion~\cite{Bluman}. By an extension of Noether's Theorem, {\em non-variational symmetries,} --- symmetries of the equations of motion which are not in general variational symmetries --- 
also lead to first integrals, which are not conservation laws of the usual divergence form, as discussed in a previous article~\cite{BludKenI}. There it was shown that a Lagrangian $\mathcal{L}(t,q_i,\dot{q_i})$ and action $S=\int{\mathcal{L}(t,q_i,\dot{q_i})} dt$, with degrees of freedom $q_i$, can be transformed under an infinitesimal point transformation
$\delta (t,q_i), \delta q_j (t,q_i)$:
\bea \delta \mathcal{L}=\dot{\mathcal{L}}\delta t+(\partial\mathcal{L}/\partial q_i)\delta q_i+(\partial\mathcal{L}/\partial\dot{q_i})
\Bigl[\frac{d \delta q_i}{d t}-\dot{q_i}\frac{d \delta t}{d t}\Bigr]=\Bigl
[\frac{dG}{dt}-
 \mathcal{L}\cdot\frac{d(\delta t)}{dt}+\mathcal{D}_i\cdot(\delta q_i-\dot{q_i}\delta t)\Bigr] \quad,
\eea
in terms of the total derivative of the {\em Noether charge,} $G:=\mathcal{L}\cdot\delta t+p_i\cdot(\delta q_i-\dot{q_i}\delta t),$ and the variational derivative $\mathcal{D}_i:= \partial\mathcal{L}/\partial q_i-d(\partial\mathcal{L}/\partial\dot{q_i})/dt$.
For transformations that leave initial and final states unchanged, the variation in action is
\be \delta S_{if}= G(f)-G(i)+\int_i^f dt\ \Bigl[\delta q_i\cdot\mathcal{D}_i+\delta t\cdot \Bigl(\frac{d \mathcal{H}}{dt}+ \frac{\partial\mathcal{L}}{\partial t}\Bigr) \Bigr]\quad , \ee
if the term in $d(\delta t)/dt$ is integrated by parts. If the system evolution obeys an action principle, that this variation vanish for independent variations $\delta q_i, \delta t$ that vanish at initial and final times, the system obeys the Euler-Lagrange equations $\mathcal{D}_i=0$ and $d \mathcal{H}/dt=-\partial\mathcal{L}/\partial t$, the rate of change of the Hamiltonian in non-conservative systems. On-shell, where $\mathcal{D}_i=0$,
\bea \delta S_{if}=\int_i^f{ \bar{\delta}\mathcal{L}}\ dt=G(f)-G(i) \\
\frac{dG}{dt} =\bar{\delta}\mathcal{L}:=\delta\mathcal{L}+\mathcal{L}\cdot (d\delta t/dt) . \eea
This is {\em Noether's equation}, giving the evolution of a symmetry generator or Noether charge, in terms of the Lagrangian transformation that it generates. It expresses the Euler-Lagrange equations of motion as the divergence of the Noether charge. This divergence vanishes for a variational symmetry, but not for any other symmetry transformation.

Noether's equation (4) could have been derived directly from the definition of the Noether charge. But using the action principle make manifest the connection between Noether's equation and the Euler-Lagrange equations. We use the action principle and this connection to reformulate the theory of hydrostatic barotropic spheres, which is integrable if they are scale symmetric,
even where this scale symmetry is not a symmetry of the action (Section II). The first integrals implied by any symmetry of the equations of motion, while generally not vanishing-divergence conservation laws, are still useful dynamical or structural first-order relationships.

Because it neglects all other structural features,
scaling symmetry is the most general simplification that one can make for any dynamical system.  For the radial scaling transformations we consider, $\delta r=r$, the Lagrangian scales as some scalar density $\delta \mathcal{L}=-2\tilde{\omega}\mathcal{L};$ and the action scales as $\delta S=(1-2\tilde{\omega}) S$.
The Noether charge generating the scale transformation evolves according to a {\em non-conservation law} $dG/dt =(1-2\tilde{\omega})\mathcal{L}$, a first-order equation encapsulating all of the consequences of scaling symmetry~\cite{BludKenI}.
From this first-order equation follow directly all the properties of index-$n$ polytropes, as established in classical works~\cite{Chandra,Schwarzschild}, modern textbooks~\cite{Kippen,Hansen}, and the recent, excellent treatments of Horedt and Liu~\cite{Horedt,Liu}.

Our secondary purpose is to present an original variational formulation of spherical hydrostatics and to extend Noether's Theorem to non-variational scaling symmetry, which yields a {\em scaling non-conservation law} (Section II). For spherical hydrostasis, we define a {\em core radius}, inside which all stars exhibit a common mass density structure.  
Outside this core, polytropes of different index $n$ show different density structures as the outer boundary is felt (Section III). Section IV completes the integration of the Lane-Emden equation by quadratures and obtains useful approximations to the Emden function $\theta_n (\xi)$.

An appendix reviews the thermodynamic properties of the physically important polytropes of index $n=3$~\cite{Hansen,Kippen,BludKenI}. What is original here is the explanation of the the differences
between relativistic degenerate white dwarf stars and ideal gas stars on the zero-age main sequence (ZAMS), following from their different entropy structures. Our original approximations to $\theta_3 (\xi)$ should prove useful in such stars.


\section{Scaling Symmetry and Integrability of Hydrostatic Spheres}  

\subsection{Variational Principle for Hydrostatic Spheres}
A non-rotating gaseous sphere in hydrostatic equilibrium obeys the equations of hydrostatic equilibrium and mass continuity
\be -d P/\rho dr= G m/r^2 , \quad d m/d r=4\pi r^2 \rho \quad, \label{eq:masspressbalance} \ee
where the pressure, mass density, and included mass $P(r),~\rho(r),~m(r)$ depend on radius $r$. For dependent variables, we use the gravitational potential $V(r)=\int_\infty^r Gm/r^2 dr$ and the thermodynamic potential (specific enthalpy, ejection energy) $H(r)=\int_{0}^{P(r)} dP/\rho $, so that~(\ref{eq:masspressbalance}) and its integrated form become
\be -d H/d r=d V/d r ,\quad V(r)+H(r)=-\frac{GM}{R}\quad, \ee
expressing  the conservation of the specific energy as the sum of gravitational and internal energies, in a star of mass $M$ and radius $R$.
The two first-order equations~(\ref{eq:masspressbalance}) are equivalent to a  second-order equation of hydrostatic equilibrium, Poisson's Law in terms of the enthalpy $H(r)$:
\be \frac{1}{r^2}\frac{d}{dr} \Big( r^2\frac{d H}{dr} \Big) + 4 \pi G \rho(H)=0 \quad , \label{eq:secondorder} \ee

We assume a chemically homogeneous spherical structure, and thermal equilibrium in each mass shell, so that $\rho(r),~P(r),~H(r)$ are even functions of the radius $r$.
At the origin, spherical symmetry requires $d P/dr=0$ and mass continuity requires, to order $r^2$,
\be \rho(r)=\rho_c (1-Ar^2), ~~m(r)=\frac{4 \pi r^3}{3}\cdot(1-\frac{3}{5}A r^2)=\frac{4 \pi r^3}{3} \cdot
\rho_c^{2/5} \rho(r) ^{3/5} \quad . \label{eq:origin} \ee
The average mass density inside radius $r$ is $\bar{\rho}(r):=m(r)/(4\pi r^3/3) = \rho_c^{2/5} \rho(r) ^{3/5}$.

In a previous paper~\cite{BludKenI}, we showed that hydrostatic equilibrium~(\ref{eq:secondorder})
follows from the variational principle $\delta W=0$ minimizing the Gibbs free energy, the integral of the Lagrangian $\mathcal{L}:$
\be W:=\int_0^R dr \mathcal{L}(r,H,H') \quad ,\ee
\be \mathcal{L}(r,H,H')=4\pi r^2[-H'^2/8 \pi G+P(\rho)] dr\quad ,\quad ':=d /dr\quad ,\ee
$W$ is the sum of the gravitational and internal specific energies per radial shell $d r$.
The canonical momentum and Hamiltonian,
\be m:=\partial\mathcal{L}/\partial H'=-r^2 H'/G\quad ,\quad
\mathcal{H}(r,H,m)=-Gm^2/2 r^2-4\pi r^2 P(H)\quad , \label{eq:hydrostaticHamilton} \ee
are the included mass and energy per mass shell.
The canonical equations are
\be \partial\mathcal{H}/\partial m=H'=-Gm/r^2\quad ,
\quad \partial\mathcal{H}/\partial H=-m'=-4\pi r^2\rho \quad .\ee
Spherical geometry makes the system nonautonomous, so that $\partial\mathcal{H}/\partial r=-\partial\mathcal{L}/\partial r=-2 \mathcal{L}/r$
vanishes only asymptotically, as the mass shells approach planarity.

The equations of hydrostatic equilibrium (5) can be rewritten
\be d\log{u}/d\log{r}=3-u(r)-n(r)v(r)\quad ,\quad
 d\log{v}/d\log{r}=u-1+v(r)-d\log{[1+n(r)]}/d\log{r}\quad, \label{eq:equil} \ee
in terms of the logarithmic derivatives
\be u(r):= d\log{m}/d\log{r}, \quad v(r):=-d\log{(P/\rho )}/d\log{r}, \quad w(r):=n(r) v(r)= -d\log{\rho}/d\log{r}\quad,\ee
and an index $n(r)$
\be n(r):=d\log{\rho}/d \log{(P/\rho)} \quad,\quad 1+\frac{1}{n (r)}:=d\log{P} /d\log{\rho}\quad ,\ee
which depends on the local thermal structure. The mass density invariant $w$ makes explicit the universal mass density structure of all stellar cores, which is not apparent in the conventional pressure invariant $v$.

\subsection{Scaling Symmetry and Reduction to First-Order Equation Between Scale Invariants}  

Following the our results~\cite{BludKenI}, a hydrostatic structure is completely integrable, if the structural equations (5) are invariant under the infinitesimal {\em scaling transformation}
\be \delta r=r,\quad \delta\rho=-n\tilde{\omega}_n\rho,\quad\delta H=-\tilde{\omega}_n H,\quad\delta H'=-(1+\tilde{\omega}_n) H',\quad \mbox{where} ~~\tilde{\omega}_n := 2/(n-1) \quad , \ee
generated by the Noether charge, for constant $n$,
\be G_n:=-\mathcal{H}\cdot r -m\cdot(\tilde{\omega}_n H)= r^2 \Bigl[ (\frac{H'^2}{2 G}+4\pi P(H))\cdot r+\tilde{\omega}_n \frac{H H'}{G} \Bigr] .\ee
The Lagrangian (10) then transforms as a scalar density of weight $-2\tilde{\omega}_n$
\be \delta\mathcal{L}=-2\tilde{\omega}_n \mathcal{L} \quad,\quad \delta S_{12}=(1-2\tilde{\omega}_n)\cdot S_{12} \quad, \ee
so that only for the $n=5$ polytrope ($\tilde{\omega}_n=1/2$) is the action invariant and scaling a symmetry of the action.

Both structural equations~(\ref{eq:equil}) are autonomous, if and only if $n$ is constant, so that,
$P(r)=K\rho(r)^{1+1/n}$, with the same constant $K$ (related to the entropy) at each radius.  When
this is so~\footnote{These characteristic equations are equivalent to a predator/prey equation in population dynamics ~\cite{Boyce,JordonSmith}. With time $t$ replacing
$-\log{r}$, they are Lotka-Volterra equations, modified by additional spontaneous growth terms $-u^2, ~w_n ^2 /n $ on the right side.
The $uw$ cross-terms lead to growth of the predator $w$ at the expense of the prey $u$, so that a population that is exclusively prey initially ($u=3,~w=0$) is ultimately devoured $u\rightarrow 0$. For the weakest
predator/prey
interaction ($n=5$), the predator takes an infinite time to reach the finite value $w_5 \rightarrow 5$.
For stronger predator/prey interaction ($n<5$), the predator grows infinitely $w_n\rightarrow\infty$ in
finite time.},
\bea du/d\log{r}=u(3-u-w_n)\quad ,\quad dw_n/d\log{r}=w_n(u-1+w_n/n) \\
\frac{d\log{w_n}}{ u-1+w_n/n }=\frac{d\log{u}}{(3-u-w_n)}=d\log{r}=\frac{d\log{m}}{u} \label{eq:chareqns12}.
\eea

In this section, we consider only the first equality in (20)
\be \frac{dw_n}{du}=\frac{w_n(u-1+w_n /n)}{u(3-u-w _n)} \quad  \label{eq:firstorder} \ee
between scale invariants, which encapsulates all the effects of scale invariance. We consider only simple polytropes with finite central density $\rho_c$, so that the regularity condition~(\ref{eq:origin}) requires that all $w _n(u)$ be tangent to $\frac{5}{3}(3-u)$ at the origin.
Such {\em Emden polytropes} are
the regular solutions $w_n(u)$ of the first-order equation (19), for which
$w _n(u)\rightarrow \frac{5}{3}(3-u)$ for $u\rightarrow 3$.

In terms of the dimensional constant, dimensional radius, and the central enthalpy and pressure
\be \alpha ^2:=\frac{(n+1)}{4\pi G}K\rho_c ^{1/n-1} ,\quad r:=\alpha\xi ,\qquad H_c:=  (n+1) (P/\rho)_{c} \equiv (n+1) K \rho^{1/n}_c ,\quad P_{c} , \ee
the second-order equation of hydrostatic equilibrium~(\ref{eq:secondorder}), takes the dimensionless form of the Lane-Emden equation
\be \frac{d}{d\xi}\Bigl( \xi^2 \frac{d\theta_n}{d\xi}\Bigr) + \xi^2\theta_n ^n = 0\quad . \label{eq:lane-emden-1} \ee
In terms of the dimensionless enthalpy $\theta_n(\xi)$ = $H/H_c$, the dimensional included mass, mass density, average included mass density, and specific gravitational force are
\be
m(r)=4\pi\rho_c\alpha^3\cdot(-\xi^2\theta_n '),~
\rho_n(r)=\rho_c\cdot\theta_n ^n(\xi),~
\bar{\rho}_n(r):=\frac{m(r)}{4\pi r^3/3}=\rho_c\cdot(-3\theta_n '/\xi),~
g(r):=4\pi\rho_c \alpha^2 (-\theta_n ')\ee
where prime designates the derivative $' :=d/d \xi$. The scale invariants are
\be u:=-\xi {\theta_n}^n/\theta_n' , \quad v_n := -\xi\theta_n '/\theta_n ,\quad \omega_n:=(u v_n^n)^{1/(n-1)}\equiv -\xi^{1+\tilde{\omega}_n} \theta_n '\quad .\ee

The Noether charge
\be G_n(\xi)=\frac{H_c ^2}{G}\cdot\Bigl\{ \xi^2 \cdot \Bigl[ \xi\Bigl(\frac{\theta_n'^2}{2}+\frac{\theta_n ^{n+1}}{n+1}\Bigr)+\tilde{\omega}_n\theta_n \theta_n '\Bigr]\Bigr\} \quad,\ee
evolves radially according to
\be \frac{d G_n}{d\xi}=(1-2\tilde{\omega}_n)\mathcal{L}=\Bigl(\frac{H_c ^2}{G}\Bigr)\cdot \Bigl(\frac{n-5}{n-1}\Bigr)\cdot\xi^2
  \Bigl( \frac{\theta_n '^2}{2}-\frac{\theta_n ^{n+1}}{n+1}\Bigr) .\ee
This non-conservation law expresses the radial evolution of energy density per mass shell, from entirely internal ($\theta_n ^{n+1}/(n+1)$) at the center, to entirely gravitational ($\theta_n '^2 /2$) at the stellar surface.

Figure~1 shows the first integrals $w_n(u)$ for $n=0, 1, 2, 3, 4, 5$. For $n=5$, scaling {\em is} a variational symmetry so that (26) reduces to a conservation law for the Noether charge
\be G_5=\frac{H_c^2}{G}\cdot \xi^2 [\xi (\frac{\theta_5 '^2}{2}+\frac{\theta_5 ^{6}}{6})+\frac{1}{2}\theta_5 \theta_5 ']=-\frac{H_c^2}{G}\cdot (uv_5 ^3)^{1/2}\cdot[-v_5-u/3 + 1]={\rm constant} \quad.\ee
For the Emden solution, $v_5$ is finite at the stellar boundary $u=0$,
the constant vanishes, and $w_5(u)=5 v_5 =\frac{5}{3} (3-u)$ everywhere.

For $n<5$, ~$v_n$ diverges at the stellar radius $\xi_1$ , but $\omega_ n\rightarrow {_0\omega_n} $, a finite constant characterizing each Emden function. At the boundary $u=0$,  our density invariant $w_n (u)$ diverges as $n [_0\omega_n ^{n-1} /u]^{1/n}$, and
\be (-\xi ^2 \theta_n ')_1= {_0\omega_n}\cdot \xi_1 ^{\frac{n-3}{n-1}} \quad. \ee

Table~I lists these constants $_0\omega_n$, along with the global mass
density ratios $\rho_{c}/\bar\rho_n(R)$ and the ensuing dimensional radius-mass relation
$M^{1-n}=[(n+1)K/G]^n \cdot (_0\omega_n ^{n-1}/4 \pi) R^{3-n}$.  Together with the well-known~\cite{Chandra,Hansen,Kippen} third, fourth and fifth columns, all of this table follows
\emph{directly} from the regular solutions of the \emph{first-order} equation~(\ref{eq:firstorder}). In addition, the sixth and seventh columns express mass concentration in an original way.

\section{Increasing Polytropic Index and Mass Concentration}  

\begin{table*}[t] 
\caption{Scaling Exponents, Core Parameters, Surface Parameters, and Mass-Radius Relations for Polytropes of Increasing Mass Concentration. Columns 3-5 are well-known~\cite{Chandra,Hansen,Kippen}. Columns 6-7 present a new measure of core concentration.}
\begin{ruledtabular}
\begin{tabular}{|l|l||l|l|l||l|r||r|}
$n$ &$\tilde{\omega}_n$ &$\xi_{1n}$ &$\rho_{cn}(R)/\bar\rho_n(R)$&$_0\omega_n$ &$r_{n{\rm core}}/R=\xi_{n{\rm core}}/\xi_1$&$m_{n{\rm core}}/M$
&Radius-Mass Relation $R^{3-n}\sim M^{1-n}/_0\omega_n$ \\
\hline 
0   &-2         &2.449              &1                  &0.333          &1         &1            &$R\sim M^{1/3}$; mass uniformly distributed \\
1   &$\pm\infty$&3.142              &3.290              &...            &0.66      &0.60         &$R$ independent of $M$ \\
1.5 &4          &3.654              &5.991              &132.4          &0.55      &0.51         &$R\sim M^{-1/3}$\\
2   &2          &4.353              &11.403             &10.50          &0.41      &0.41         &                     \\
3   &1          &6.897              &54.183             &2.018          &0.24      &0.31         &$M$ independent of $R$ \\
4   &2/3        &14.972             &622.408            &0.729          &0.13      &0.24         &                      \\
4.5 &4/7        &31.836             &6189.47            &0.394          &0.08      &0.22         &                 \\
5   &1/2        &$\infty$           &$\infty$           &0              &0         &0.19         &$R=\infty$ for any $M$; mass infinitely concentrated \\
\end{tabular}
\end{ruledtabular}
\end{table*}

{\em Emden functions} are the normalized regular solutions of the Lane-Emden equation~(\ref{eq:lane-emden-1}) for which the mass density is finite at 
the origin, so that $\theta_n (0)=1,~\theta_n '(0)=0$. Each Emden function of index $n$
is characterized by its first zero $\theta_n (\xi_{1n})=0$, at dimensionless boundary radius $\xi_{1n}$.
As an alternative measure of core concentration, we define the {\em core radius} $\xi_{\rm core}$ implicitly by $u(\xi_{\rm core}):=2$,
where gravitational and pressure gradient forces are maximal.
This core radius, where $w_n\approx 2$ and the mass density has fallen to $\rho_{n{\rm core}}/\rho _{nc}\approx 0.4$ for all polytropes $n\geq 1$, is
marked by red dots in Figures 1, 2, 3.
The sixth and seventh columns in Table~I list dimensionless values for the fractional core radius $r_{n{\rm core}}/R=\xi_{n{\rm core}}/\xi_1$ and fractional included mass $m_{n{\rm core}}/M$.  Within the core $u>2$, the internal energy dominates over the gravitational energy, so that for $n\geq 1$,
\be w_n(u) \approx w_5(u)=\frac{5}{3} (3-u)\quad,\quad \theta_n (\xi)\approx 1-\xi^2 /6 \quad,\quad \mbox{for}\quad u_n>2,~\xi<\xi_{core} \label{eq:coredef} \quad ,
 \ee
consistent with the universal density structure (8) all stars enjoy near their center.

\begin{figure}[b]
\includegraphics[scale=0.95]{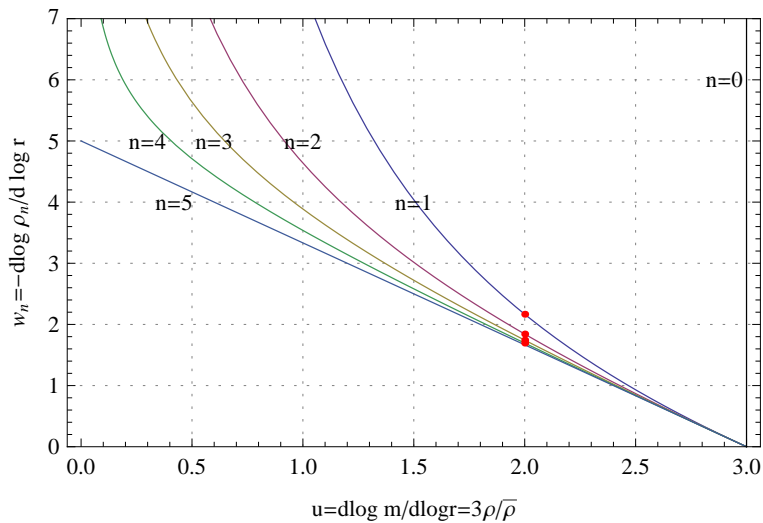} 
\caption{Dilution of polytrope mass density as the boundary is approached ($u\rightarrow 0$). All solutions are tangent to the same density structure $w_n(z)\rightarrow w_5=(5/3)(3-u)$ at the center ($u=3$), but differ for $u<2$ outside the core. Approaching the outer boundary ($u_n\rightarrow 0$), 
the density $\rho_n(r)$ falls rapidly, but $u v_n ^n:=\omega_n ^{n-1}$ approaches a constant $_0\omega_n ^{n-1}$ so that $w_n \rightarrow n[_0\omega_n ^{n-1}/u_n]^{1/n}$ diverges, for $n<5$.}
\end{figure}

\begin{description}    
\item[For $n=0$,] the mass is uniformly distributed, and the entire star is core.
\item[As $0<n<5$ increases,] the radial distribution concentrates, and the envelope outside the core grows. With increasing core concentration:
    \begin{description}
    \item [For $1<n<3$,] the radius $R$ decreases with mass $M$. Nonrelativistic degenerate stars have $n=3/2$.
    \item [For n=3,] the radius $R$ is independent of mass $M$.  This astrophysically important case is discussed in Section IV and the Appendix.
    \item [For $n>3$,] the radius $R$ increases with mass $M$.
    As $n \rightarrow 5$, the stellar radius increases $\xi_{1n} \rightarrow 3(n+1)/(5-n)$, the core radius shrinks $\xi_{\rm core} \rightarrow \sqrt{10/3 n}$, the fractional core radius $r_{\rm core}/R=\xi_{\rm core}/\xi_{1n} \rightarrow 0.045(5-n)$, $m_{n{\rm core}}/M \rightarrow 0.19$, and $_0\omega_n \rightarrow \sqrt{3/\xi_{1n}}\rightarrow 0$.
    \item[For $n=5$,] the mass is infinitely concentrated toward the center, and the stellar radius $R=\infty$ for any mass $M$. Scaling becomes a variational symmetry, so that the Noether charge $G_5$ in (40) is constant with radius. For the regular solution this constant vanishes:
    \be G_5 \sim [\xi (\frac{\theta_5'^2}{2}+\frac{\theta_5 ^{6}}{6})+\frac{1}{2}\theta_5 \theta_5 ']=(uv_5 ^3)^{1/2}\cdot(v_5-u/3-1)=0 \quad, \ee
    so that $v_5=1-u/3,~\theta_5 '=-\xi \theta_5 ^3/3$.
    Integrating then yields
    \be \theta_5(\xi)=(1+\xi^2 /3)^{-1/2}\quad , \ee
    after normalizing to $\theta_5 (0)=1$.
    \item[For $n>5$] the central density diverges, so that the total mass $M$ is infinite.
    \end{description}
\end{description}

\section{Regular Emden Solutions and Their Approximations} 

In place of $u$, we now introduce an equivalent homology invariant
$z:=3-u=-d\log{\bar{\rho}_n}/d\log{r}$, where $\bar{\rho}_n:=3 m (r)/4\pi r^3$ is the average mass density inside radius $r$.
In term of $z,~w_n$, the characteristic differential equations~(\ref{eq:chareqns12}) are
\be \frac{d z}{(3-z)(w_n-z)}=\frac{d\log{w_n}}{2-z+w_n/n}=d \log{r}=\frac{d\log{m}}{3-z}\quad .\ee

\begin{figure}[t] 
\includegraphics[scale=0.75]{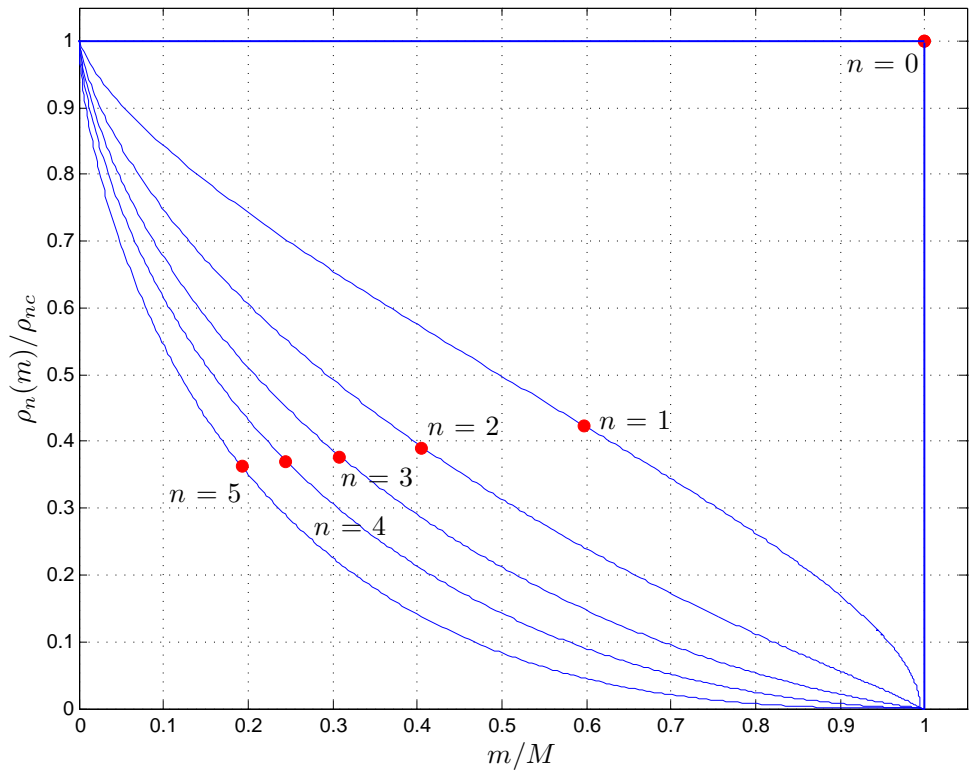}
\caption{Normalized mass density profiles as a function of fractional included mass $m/M$, for polytropes of mass concentration increasing with $n$. The red dots mark the core radii, at which the densities stay near
$\rho(r_{\rm core})/\rho_c \approx 0.4$, for all $n \geq 1$. For uniformly distributed mass ($n=0$), the polytrope
is all core. As the mass concentration increases ($n\rightarrow 5$), the core shrinks to about 20\% of the mass.}
\end{figure}

\begin{figure}[t] 
\includegraphics[scale=0.75]{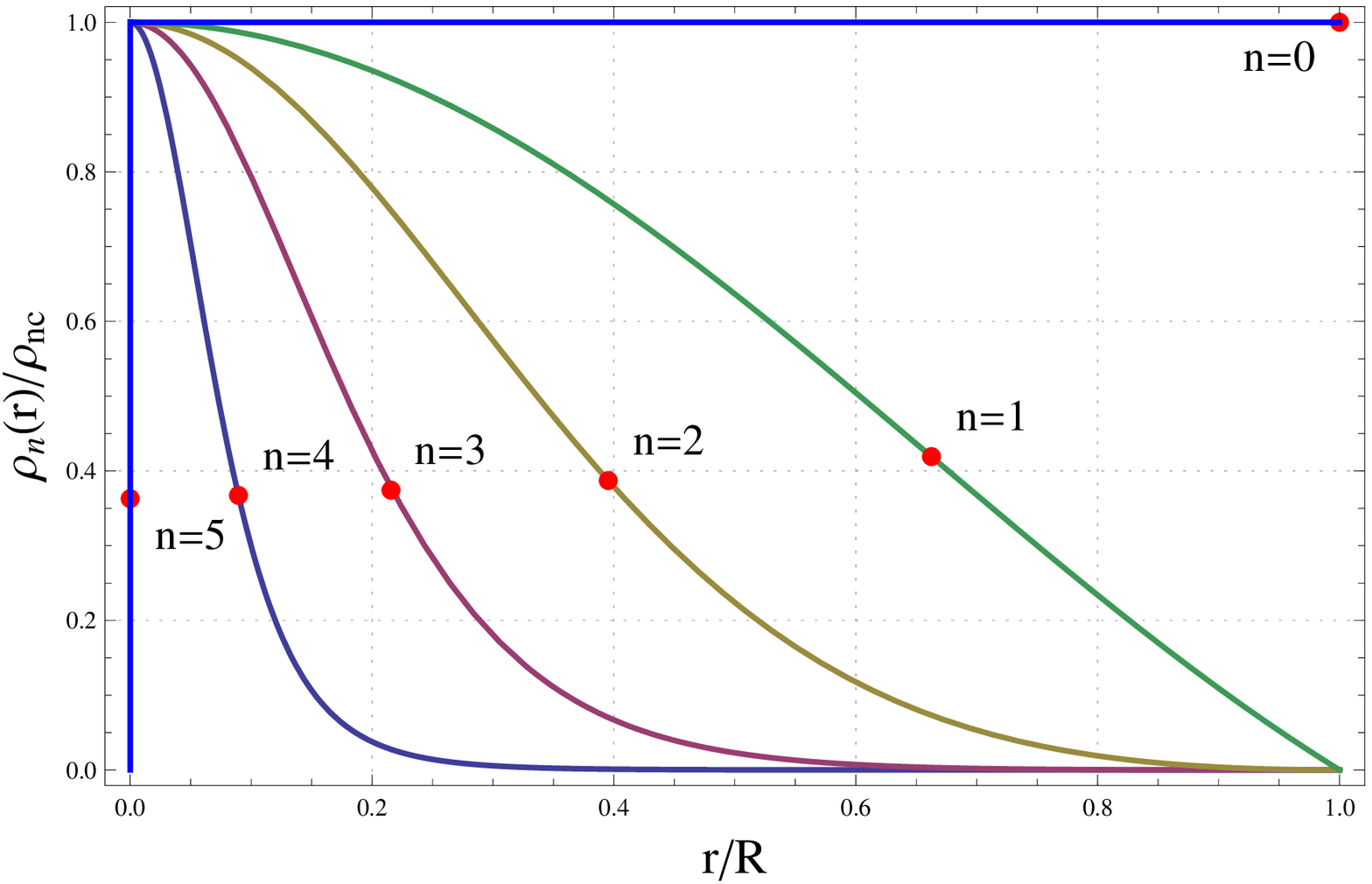}
\caption{Normalized mass density profiles as function of fractional radius $r/R$. The density is uniform for $n=0$, but is maximally concentrated at finite radius for the $n=5$ polytrope, which is unbounded ($R=\infty$). The density at the core radius stays about $\rho(r_{\rm core})/\rho_c \sim 0.4$, for any $n\geq 1$.}
\end{figure}

Incorporating the boundary condition, the first of equations (40) takes the form of a Volterra integral equation~\cite{BludKen}
\be w_n (z)=\int_0 ^z\ dz\ w_n\frac{(2-z+w_n /n)}{(3-z)(w_n-z)}\approx (5/J_n)[1-(1-z/3)^{J_n}]:=w_{n{\rm Pic}}(z)\quad ,\quad J_n := (9n-10)/(7-n) .\ee
The {\em Picard approximation} is defined by inserting the core values $w_n (z)\approx$
$(5/3)z$ inside the preceding integral. For $n=0,~5$, this Picard approximation is everywhere exact. For intermediate polytropic indices $0<n<5$, the Picard approximation breaks down
approaching the boundary, where $w_n$ diverges as $w_n \rightarrow n[_0\omega_n ^{n-1}/u]^{1/n}$, and is poorest for $n\approx 3$.

\begin{table*}[t] 
\caption{Taylor Series and Picard Approximations $\theta_{n{\rm Pic}}(\xi)$ to Emden Functions $\theta_{n}(\xi)$}
\begin{ruledtabular}
\begin{tabular}{|l||l||l|l}
$n$  &Emden Function $\theta_n(\xi)$ and Taylor Series &$N_n:=5/(3n-5)$ &Picard Approximation $\theta_{n{\rm Pic}}(\xi):=
(1+\xi^2/6N_n)^{-N_n}$ \\
\hline \hline
0    &$1-\xi^2/6$                                   &-1         &$1-\xi^2/6$                              \\
1    &$\sin{\xi}/\xi=1-\xi^2/6+\xi^4/120-\xi^6/5040+\cdots$&-5/2&$(1-\xi^2 /15)^{5/2}=1-\xi^2/6+\xi^4/120-\xi^6/10800+\cdots$    \\
$n$  &$1-\xi^2/6+n \xi^4/120-n(8n-5)/15120 \xi^6+\cdots$   &$5/(3n-5)$ &$(1+\xi^2/6N_n)^{-N_n}=1-\xi^2/6+n \xi^4/120-n(6n-5) \xi^6 /10800+\cdots$ \\
5    &$(1+\xi^2/3)^{-1/2}$                                 &1/2        &$(1+\xi^2/3)^{-1/2}$ \\
\end{tabular}
\end{ruledtabular}
\end{table*}

After obtaining $w_n(z):=-d\log{\rho_n}/d\log{r}$, either numerically or by Picard approximation, another integration gives~\cite{BludKen}
\bea
\rho_n(z)/\rho_{cn} =\exp{\Bigl\lbrace -\int_0 ^z  \frac{dz\ w_n(z)}{[w_n(z)-z](3-z)} \Bigr\rbrace} \approx(1-z/3)^{5/2} \\
\theta_n =[\rho_n(z)/\rho_{cn}]^{1/n} = \exp{\Bigl\lbrace  -\int_0 ^z  \frac{dz\ w_n(z)}{n [w_n(z)-z](3-z)}\Bigr\rbrace}\approx  (1-z/3)^{5/2n}:=\theta_{n{\rm Pic}} \\
m(z)/M=(\frac{z}{3})^{3/2}\cdot \exp{\Bigl\lbrace \int_3 ^z dz \Bigl\lbrace
\frac{1}{[w_n(z)-z]}-\frac{3}{2z}\Bigr\rbrace \Bigr\rbrace} \approx \Bigl(\frac{z}{3}\Bigr)^{3/2}\\
r(z)/R=\xi /\xi_{1n}=(\frac{z}{3})^{1/2}\cdot \exp{\Bigl\lbrace  \int_3 ^z  dz \Bigl\lbrace \frac{1}{(3-z)[w_n(z)-z]}-
\frac{1}{2z}\Bigr\rbrace \Bigr\rbrace } \approx \frac{(3z)^{1/2}}{3-z}\quad .  \eea
All the scale dependance now appears in the integration constants $M$ and $R(M)$, which except for $n=3$  depends on $M$.
Inserting the core values $w_n (z)\approx$
$(5/3)z$ inside the integral, the Picard approximations
\be \theta_{n{\rm Pic}}(\xi)=(1+\xi^2/6N_n)^{-N_n}\quad, \quad N_n:=5/(3n-5) \ee
to the Emden functions are obtained and tabulated
in the last column of Table~II. For polytropic indices $n=0,~5$, this Picard form is exact.
For intermediate polytropic indices $0<n<5$, the Picard approximation remains a good approximation through order $\xi^6$, but breaks down approaching the outer boundary. Unfortunately, the Picard approximation is poorest near $n=3$, the astrophysically most important polytrope.  

Figure 4 compares three approximations to this most important Emden function, shown in yellow, whose Taylor series expansion is
\be \theta_3(\xi)= 1-\xi^2/6+\xi^4 /40-(19/5040) \xi^6+(619/1088640) \xi^8 -(2743/39916800) \xi^{10} + \cdots\quad . \ee

\begin{description}
\item[Tenth-order polynomial approximation] to this Taylor series expansion
\be 1-0.1666667 \xi^2+ 0.025 \xi^4 - 0.0037698 \xi^6 +  0.0005686 \xi^8-0.00006872 \xi^{10} \quad,  \ee
shown in red, diverges badly for $\xi >2.5\approx 1.7~\xi_{3{\rm core}}$.
\item[Picard approximation]
\bea \theta_{3{\rm Pic}}(\xi)=(1+2\xi^2 /15)^{-5/4}=1-\xi^2 /6+\xi^4 /40 -13\xi^6 /3600 + \cdots \\
=1-0.1666667 \xi^2+0.025 \xi^4-0.003611 \xi^6 +\cdots,
\eea
shown in dashed green, converges and remains a good approximation over the bulk of the star, with $\leq 10\%$ error out to $\xi\approx 3.9$, more than twice the core radius and more than half-way out to the stellar boundary at $\xi_{13}=6.897$. This approximation suffices in white dwarf and ZAMS stars, except for their outer envelopes, which are never polytropic and contain little mass.
Because it satisfies the central boundary condition, but not the outer boundary condition, the Picard approximation underestimates $\theta ' (\xi)$ and overestimates $\theta (\xi)$ outside $\xi \sim 3.9$.
\item[Pad\'{e} rational approximation]~\cite{PadeApprox,Seidov}:
\be \theta_{\rm 3Pad}=
\frac{1-\xi^2/108+11 \xi^4/45360}{1+17\xi^2/108+ \xi^4/1008}=1- 0.166667 \xi^2 + 0.025 \xi^4 - 0.00376984 \xi^6 +
 0.0005686 \xi^8 - 0.0000857618 \xi^{10}+\cdots , \ee
shown in dashed heavy black, is a simpler and much better approximation. By construction, it agrees with
the series expansion~(40) through fourth order. In fact, this Pad\'{e} approximation is almost exact out to its first zero at
$\xi_1=6.921$, very close to the true outer boundary $\xi_{13}=6.897$.
\end{description}
These simple analytic approximations to $\theta_3 (\xi)$ simplify structural modeling of massive white dwarfs and ZAMS stars.

\begin{figure}[t] 
\includegraphics[scale=0.7]{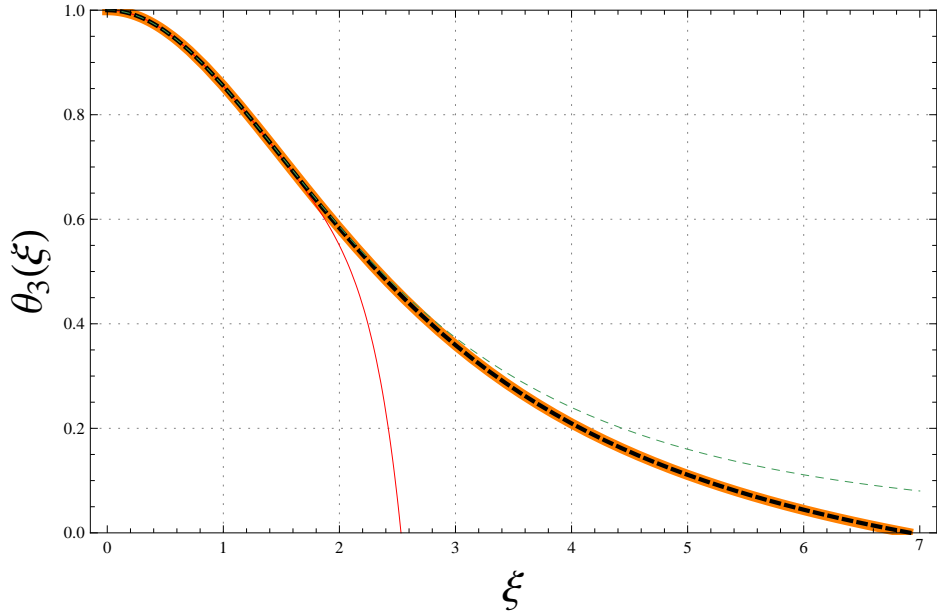}
\caption{The exact Emden function $\theta_3(\xi)$ (solid yellow) and its polynomial (red), Picard (green dashed) and Pad\'{e} (heavy black dashed)
approximations. Even in this worst case, the Picard approximation holds out to twice the core radius
at $2 \xi_{3{\rm core}}= 3.3$, before breaking down near the boundary. The Pad\'{e} approximation is indistinguishable from the exact solution, vanishing at $\xi_1=6.921$, very close to the true boundary at $\xi_{13}=6.897$.}
\end{figure}

\section{Conclusions} 

We have explored how a symmetry of the equations of motion, but not of the action, reduces a second-order differential equation to first-order, which can be integrated by quadrature. In scale-invariant hydrostatics, the symmetry of the equations yields a first integral, which is a first-order equation between scale invariants, and yields {\em directly} all the familiar properties of polytropes.

We observe that, like all stars, polytropes of index $n$ share a common core density profile and defined a {\em core radius} outside of which their envelopes differ. The Emden functions $\theta_n(\xi)$, solutions of the Lane-Emden equation that are regular at the origin, are finally obtained, along with useful approximations.

The Appendix reviews the astrophysically most important $n=3$ polytrope, describing relativistic white dwarf stars and zero age main sequence stars. While reviewing these well-known applications~\cite{Hansen,Kippen}, we stress how these {\em same} mechanical structures differ {\em thermodynamically} and the usefulness of our original (Section IV) approximations to these Emden functions.

\appendix*

\section*{Appendix: Astrophysical Applications of the $n=3$ Polytrope} 

The $n=3$ polytrope, which is realized in white dwarfs of maximum mass and in the Eddington standard model for ZAMS stars just starting hydrogen burning, is distinguished by a unique $M-R$ relation: the mass $M=4\pi (_0\omega_3) ~(K/\pi G)^{3/2}$ is independent of radius $R$, but depends on
the constant $K:=P/\rho^{4/3}$.  In these stars, the gravitational and internal energies cancel, making the 
total energy $W=\Omega + U=0$. Because these stars are in neutral mechanical equilibrium at any radius, they can expand or contract homologously.

\subsection{Relativistic Degenerate Stars: $K$ Fixed by Fundamental Constants} 

The most massive white dwarfs are supported by the degeneracy pressure of relativistic electrons, with number density $n_e=\rho/\mu_e m_H$, where $m_H$ is the atomic mass unit and the number of electrons per atom $\mu_e=Z/A=2$, because these white dwarfs are composed of pure He or ${\rm C}^{12}/{\rm O}^{16}$ mixtures. Thus, $K_{WD}=(hc/8) [3/\pi ]^{1/3} {m_H \mu_e}^{-4/3}$ depends only on fundamental constants. This universal value of $K_{WD}$ leads to the limiting Chandrasekhar mass $M_{\rm Ch}=(\pi^2/8\sqrt{15}) M_{\star}/\mu_e^2=5.824 M_{\odot}/\mu_e ^2=1.456 M_{\odot}\cdot(2/\mu )^2$~~\cite{Hansen,Kippen}.

\subsection{Zero-Age Main Sequence Stars: Mass and $K(M)$ Dependent on Specific Radiation Entropy} 

In an ideal gas supported by both gas pressure $P_{\rm gas}=\mathcal{R}\rho T/\mu :=\beta P$ and radiation pressure $P_{\rm rad}=a T^4 /3 :=(1-\beta) P$, the radiation/gas pressure ratio is
\be \frac{P_{\rm rad}}{P_{\rm gas}}:=\frac{1-\beta}{\beta}=\frac{T^3}{\rho}\cdot\frac{a \mu}{3 \mathcal{R}}\quad.\ee
The specific radiation and ideal monatomic gas entropies are
\be S_{\rm rad}=\frac{4a T^3}{3 \rho}, \quad \quad S_{\rm gas}(r)=
\Bigl(\frac{\mathcal{R}}{\mu}\Bigr)\cdot \log{\Bigl[\frac{T(r) ^{5/2}}{\rho (r)}\Bigr]}\quad  ,\ee
so that the gas entropy gradient
\be \frac{d S_{\rm gas}}{d\log{P}}=\Bigl(\frac{5\mathcal{R}}{2\mu}\Bigr)\cdot (\nabla -\nabla_{ad})=\Bigl(\frac{\mathcal{R}}{\mu}\Bigr) \cdot \Bigl(\frac{\nabla}{\nabla_{ad}} -1\Bigr) \ee
depends on the difference between the adiabatic gradient $\nabla_{ad}=2/5$ and the star's actual thermal gradient $\nabla:=d\log{T}/d\log{P}$, which depends on the radiation transport.

Bound in a polytrope of order $n$ , the ideal gas thermal gradient and gas entropy gradient are
\be \nabla:=1/(n+1) \quad,\quad \frac{d S_{\rm gas}}{d\log{P}}=\Bigl(\frac{\mathcal{R}}{\mu}\Bigr)\cdot \Bigl[\frac{5}{2(n+1)}-1\Bigr]\quad.\ee
For $n>3/2$, the thermal gradient is subadiabatic, the star's entropy increases outwards, so that the star is stable against convection.

ZAMS stars, with mass $0.4 M_{\bigodot} <M<150 M_{\bigodot}$, have nearly constant radiation entropy $S_{\rm rad}(M)$, because radiation transport leaves the luminosity generated by interior nuclear burning everywhere proportional
to the local transparency (inverse opacity) $\kappa ^{-1}$.  Assuming constant  $S_{\rm rad} (M)$, we have {\em Eddington's standard model}, an $n=3$ polytrope with $S_{\rm rad}(M)=
4 (\mathcal{R}/\mu)\cdot(1-\beta)/\beta$ and
\be K(M)=P/\rho^{4/3}=\lbrace [3(1-\beta)/a] (\mathcal{R}/\mu\beta )^4 \rbrace^{1/3} ,  \ee
depends only on $\beta (M)$, which is itself determined by
{\em Eddington's quartic equation}~\cite{Kippen,Hansen,Chandra}
\be \frac{1-\beta}{\beta ^4}=\Bigl( \frac{M\mu^2}{M_{\star}} \Bigr)^2\quad ,\quad M_{\star}:=\frac{3\sqrt{10} ~_0\omega_3}{\pi ^3} \Bigl( \frac{hc}{G m_H^{4/3}} \Bigr)^{3/2}=18.3 M_{\odot}\quad .\ee

The luminosity
\be L=L_{\rm Edd} [1-\beta (M)]=L_{\rm Edd}\cdot (0.003)\mu ^4 \beta (M)^4 (M/M_\odot)^3 ,\ee
depends on the {\em Eddington luminosity} $L_{\rm Edd}:=4\pi c G M/\kappa_p$ through the photospheric opacity $\kappa_p$.
This mass-luminosity relation is confirmed in ZAMS stars: on the lower-mass ZAMS, $\beta \approx 1,~L\sim M^3$; on the upper-mass ZAMS, $\beta\approx (M\mu^2/M_{\star})^{-2} \ll 1,~L\sim M $~\cite{Hansen}.

\begin{acknowledgments}
Thanks to Andr${\rm\acute{e}}$s E. Guzm${\rm\acute{a}}$n (Universidad de Chile) for calculating the figures with Mathematica and proofreading the manuscript. SAB was supported by the Millennium Center for Supernova Science through grant P06-045-F funded by Programa Bicentenario de Ciencia y Tecnolog\'ia de CONICYT and Programa Iniciativa Cient\'ifica Milenio de MIDEPLAN.
The referees improved the form and focus of the paper.
\end{acknowledgments}

\bibliography{bibliographyLE}
\end{document}